\documentclass[AMA,STIX1COL]{WileyNJD-v2}

\articletype{Article Type}%

\received{26 April 2016}
\revised{TBA}
\accepted{TBA}

\usepackage{bbm}
\usepackage{enumerate}
\usepackage{url}            
\usepackage{threeparttable} 
\usepackage{adjustbox}      
\usepackage{longtable}      
\usepackage{setspace}       
\usepackage{xcolor}         
\usepackage{pbox}           

\graphicspath{{figures/}}

\begin{document}

\title{Number of Repetitions in Re-randomization Tests\protect\thanks{The authors would like to thank Dr. Olga Kuznetsova, Dr. Keaven Anderson, Dr. Gregory Golm, and Dr. Yue Shentu  for helpful comments and suggestions. }}

\author[1]{Yilong Zhang}

\author[2]{Yujie Zhao}

\author[2]{Yiwen Luo}

\authormark{Yilong Zhang \textsc{et al}}

\address[1]{\orgdiv{Reality Lab}, \orgname{Meta Platforms, Inc.}, \orgaddress{\state{CA}, \country{USA}}}

\address[2]{\orgdiv{Biostatistics and Research Decision Sciences }, \orgname{Merck \& Co., Inc}, \orgaddress{\state{NJ}, \country{USA}}}

\corres{*Yilong Zhang \\ \email{elong0527@gmail.com}}


\abstract[Abstract]{In covariate-adaptive or response-adaptive randomization, the treatment assignment and outcome can be correlated. Under this situation, re-randomization tests are a straightforward and attractive method to provide valid statistical inference. In this paper, we investigate the number of repetitions in the re-randomization tests. This is motivated by the group sequential design in clinical trials, where the nominal significance bound can be very small at an interim analysis. Accordingly, re-randomization tests lead to a very large number of required repetitions, which may be computationally intractable. To reduce the number of repetitions, we propose an adaptive procedure and compare it with multiple approaches under pre-defined criteria. Monte Carlo simulations are conducted to show the performance of different approaches in a limited sample size. We also suggest strategies to reduce total computation time and provide practical guidance in preparing, executing and reporting before and after data are unblinded at an interim analysis, so one can complete the computation within a reasonable time frame. }


\jnlcitation{\cname{%
\author{Y Zhang},
\author{Y Zhao}, and
\author{Y Luo}} (\cyear{2021}),
\ctitle{Number of Repetitions in Re-randomization Tests}, \cjournal{under the review of Statistics in Medicine}, \cvol{2021+}.}

\maketitle


\section{Introduction}
\label{sec: introduction}
In covariate-adaptive or response-adaptive randomizations, it is possible that the treatment assignment and outcome can be correlated. 
To overcome this type of correlation, one can use the re-randomization test by fixing the order of randomization of study subjects, their covariates, and/or responses.
The aforementioned re-randomization tests are a straightforward and attractive method to provide valid statistical inference, especially under the covariate-adaptive or response-adaptive randomization 
\citep{rosenberger2015randomization}.
One key feature of re-randomization tests is the sufficient number of repetitions (denoted as $L$) to estimate p-values. 
In the existing literature, one of the popular way\citep{plamadeala2012sequential} to select $L$ is to ensure 
\begin{equation*}
    P(|\hat{p} - p| \le 0.1 p) = 0.99,    
\end{equation*}
where $p$ denotes the true p-value and $\hat{p}$ is an estimator of $p$. 
The above equation implies 
\begin{equation}
    L
    \approx 
    (2.576/0.1)^2 (1 - p) / p.
\label{eq:l1}
\end{equation}
Formula \eqref{eq:l1} works well when $p \geq 0.01$.
For example, when $p = 0.01$, we have $L \approx 65,695$, which is not dramatically large. 
So it does not cost much time to finish the $65,695$ repetitions.
However, $L$ can be computationally intractable to estimate a small $p$. 
For example, $L \approx 6,635,113$ is needed to estimate a true p-value $p = 0.0001$ (see detailed introduction of \eqref{eq:l1} is available in Section \ref{sec: review of existing methods}).

In practice, it is common to have a tiny nominal significance bound $\alpha$ in the covariate-adaptive randomization, especially when one calculates the efficacy boundary in interim analyses for clinical trials based on group sequential design.  
One motivating example is a hypothetical oncology trial with two co-primary endpoints, i.e., progression-free survival (PFS) and overall survival (OS).
For this trial, we use a group sequential design with one interim analysis.
The initial one-sided $\alpha-$level for testing PFS and OS can be set at $0.005$ and $0.020$ respectively for an overall one-sided $\alpha-$level at $0.025$ based on the graphical method \citep{maurer2013multiple} to control multiplicity for multiple hypotheses as well as interim analyses. 
For PFS, the efficacy boundary is approximately $0.000072$ 
for the first interim analysis using a Lan-DeMets O'Brien-Fleming spending function. 
In the Appendix, we provide R code to reproduce this group sequential design that requires approximately 744 participants.

In the above motivating example, we also assume the Pocock and Simon covariate-adaptive allocation procedure \citep{pocock1975sequential} is used to provide balance in intervention assignments within the categories of a list of factors . 
The allocation procedure is commonly referred to as the minimization randomization procedure \citep{jin2019algorithms}. 
In this motivating trial, three randomization stratification factors are considered: (1) 50 study sites, (2) Eastern Cooperative Oncology Groups (ECOG) performance status, and (3) tumor mutation burdens (TMB), which gives up to 300 randomization strata. 
After participants are randomized using minimization procedure, a rerandomization stratified log-rank test is preferred to analyze PFS or OS. 

For this motivating clinical trial, one challenging issue is to determine the number of repetitions $L$, if a re-randomization test is desired based on a stratified log-rank test to evaluate the hypothesis of a treatment difference in PFS.
By applying the efficacy boundary at $0.000072$ in the first interim analysis using formula \eqref{eq:l1}, the analysis requires repeating around 9.2 million times for approximately 650 participants at the first interim analysis. 

In practice, it can be a challenge to complete 9.2 million repetitions within a time-sensitive period after unblinding the data at an interim analysis. 
The computation issue becomes even more challenging if the re-randomization test is required to estimate the hazard ratio and its 95\% confidence interval from a stratified Cox model. 
The computation issue may also apply to any sensitivity analyses that will be performed, where the same procedure with different censoring rules is repeated.
However, intuitively, one can reduce the number of repetitions if it is a testing problem to report a p-value whose exact value is far from the pre-defined nominal significance bound. 
For example, if the treatment effect is not statistically significant and the estimated p-value after unblinding data is 0.1 based on 10,000 repetitions, it may be safely concluded that the study needs to continue based on the efficacy boundary at $0.000072$ without 9.2 million repetitions. 

In this paper, we propose an adaptive procedure with the objective of reducing the number of repetitions, and we compare multiple approaches under pre-defined criteria.
We also suggest strategies to reduce total computation time and provide practical guidance in preparing, executing, and reporting before and after data are unblinded at an interim analysis to complete the computation within a limited time frame. 

The rest of this paper is organized as follows. 
Section \ref{sec: methods} proposes multiple approaches to pre-define the number of repetitions for the re-randomization test. 
Section \ref{sec: simulation} presents Monte Carlo simulation results to compare different proposed approaches. 
Section \ref{sec: computation time} discusses the strategies to reduce total computation time.
Finally, Section \ref{sec: discussion} discusses practical guidance to practitioners in preparing an interim analysis before and after database lock.

\section{Methods}
\label{sec: methods}
The problem of choosing the number of repetitions has been investigated in relevant areas such as 
Bootstrap \citep{davidson2000bootstrap, andrews2000three} and 
Markov Chain Monte Carlo (MCMC) \citep{gelman1992inference, roy2020convergence}. 
Beyond these two areas, our motivating problem comes from 
the re-randomization test under a given nominal significance bound $\alpha$.
The same framework can be used to determine the number of repetitions for the confidence interval and confidence regions, as discussed by \cite{andrews2000three}.
Following this logic, one can calculate the number of repetitions required for PFS in our motivating example, which would be the same as that required for a stratified log-rank test and 95\% confidence interval estimated from a stratified Cox model.

In hypothesis testing, the quantities $(p, \hat p_{\infty}, \hat p_L)$ of interest are the ``exact'' p-value (ground truth), the ``ideal'' rerandomized p-value based on infinite repetitions, and the ``real-practice'' reranomized p-value based on $L$ repetitions.
We notice that $\hat p_{\infty}$ has the desirable property that it eliminates numeric error from a finite re-randomization procedure. 
Yet, in reality, it is impossible to obtain $\hat{p}_\infty$. 
We seek to find a proper $L$ such that the estimated p-value $\hat{p}_L$ can draw consistent inference conclusions as 
$\hat{p}_\infty$ by comparing the nominal significance bound $\alpha$.

In this paper, we considered only one-sided tests without loss of generality. 
For a two-sided test, one can simply replace $\alpha$ with $\alpha/2$.

\subsection{Review of Existing Methods}
\label{sec: review of existing methods}
\begin{itemize}
    
    \item[M1] \textbf{PR Re-randomization}. 
    \cite{plamadeala2012sequential} propose a simple formula to control the error rate.
    We denote this method as \textit{PR re-randomization}.  
    Specifically, this approach selects the number of repetitions $L$ to quantify the error rate between $\hat{p}_L$ (the ``real-practice'' p-value estimated from $L$ repetitions) and $p$ (the ``exact'' p-value) by the following criterion  
    \begin{equation*}
      L = \{L: P(|\hat{p}_L - p| \le 0.1 p) = 0.99\}, 
    \end{equation*}
    Equivalently, we have  
    \begin{equation*}
      L \approx (2.576/0.1)^2 (1 - p) / p,
    \end{equation*}
    where $2.576$ is the numerical value of $\Phi^{-1}(0.995)$ with $\Phi^{-1}(\cdot)$ as the quantile function for the standard normal distribution.
    Heuristically, this is a direct result from
    \begin{equation}
      \hat{p}_{L} 
      \sim 
      \text{Binomial}(L, p) \approx_F \mathcal{N}(p, p(1-p) / L),
    \label{eq:p-asy}
    \end{equation}
    when $L$ is large enough.
    In practice, the true p-value $p$ is unknown.
    So, we may use the nominal significance bound $\alpha$ to replace $p$ in hypothesis testing problems and determine $L$ by
    \begin{equation}
    \label{equ: PR formula}
      L 
      \approx 
      (2.576/0.1)^2 (1 - \alpha) / \alpha.
    \end{equation}
    
    One limitation of the \cite{plamadeala2012sequential} method is that the number of repetitions $L$ becomes impractical when $\alpha$ becomes very small, which is possible in an interim analysis of a group sequential design as discussed in the previous section. 

   \item[M2] \textbf{Parametric Re-randomization}. 
   This method is inspired by the parametric bootstrap procedure \citep{efron1994introduction} to reduce the number of repetitions.
   We denote this method as \textit{parametric re-randomization} in this paper. It assumes the test statistics asymptotically follow a known distribution with unknown parameters under the null hypothesis, for example, a normal distribution with unknown mean and standard deviation. 
   Given the known type of distribution, we can first estimate its unknown distribution parameters, which may require a much smaller number of repetitions to control the numeric error rate \citep{efron1990more}.  
   Then, we can draw statistical inference based on the $(1-\alpha)$\textsuperscript{th} quantile of the estimated distribution.
   Although the idea of parametric re-randomization is attractive, the validity of this procedure is unknown for covariate-adaptive randomization. Because the samples are not i.i.d. and the asymptotic distribution of the test statistics are commonly unknown in theory. 
   In the framework of the parametric bootstrap, \cite{efron1990more} suggested 50-200 repetitions to estimate the various percentiles.
   For parametric re-randomization, the required number of repetitions can be larger.
   In Section 3, we investigate the numerical performance through simulation with different numbers of repetitions to understand the potential impact when the nominal significance bound  $\alpha$ is very small. 

\end{itemize}


\subsection{Adaptive Procedure} 

Under the hypothesis testing framework, the ``exact'' p-value $p$ is unknown.
However, one can always estimate the p-value through a certain number of repetitions and get an estimation $\hat p_{L}$. 


Our proposed method is an adaptive procedure, which continuously monitors the estimated p-value $\hat p_{L}$ under a sequence of repetitions $L$.
For example, suppose the sequence of repetitions is $L \in \{5,000, 15,000, 25,000, \ldots\}$, and the monitored p-value will be $\hat p_{5,000}, \hat p_{15,000}, \hat p_{25,000}, \ldots$. 
If $\hat p_{5,000} = 0.1$, we may safely draw a conclusion
for a nominal significance bound at $\alpha = 0.0001$. 

The key to the above adaptive procedure is to define a proper stopping rule to guide it in a sequence of repetitions.
Let $m_L$ denote the number of ``events'' -- defined as the number of test statistics with more extreme values compared with observed test statistics.
Accordingly, one can estimate the p-value as $\hat p_L = m_L/L$.
To ensure a justified $\hat p_L$, 
we proposed to control both the upper bound and the lower bound, i.e.,
\begin{align}
    P\left( \hat{p}_L  > (1 + \delta_u) \alpha \right) & = \rho_u  \label{equ: upper bound}    \\
    P\left( \hat{p}_L  < (1 - \delta_l) \alpha \right) & = \rho_l.
    \label{equ: lower bound}
\end{align}
Here $\delta_u, \delta_l > 0$ and $\rho_u, \rho_l \in (0, 1)$ is the parameters pre-specified by users.
And they jointly ensure that the estimation $\hat{p}_L$ is not too away from the ground truth $\alpha$ by a large probability.
For example, one can set $\delta_u = \delta_l = 0.1$ and $\rho_u = \rho_l = 0.99$.
In this way, the estimation $\hat{p}_L$ is controlled within $(0.9\alpha, 1.1\alpha)$ with a probability of $0.99$.

By using the distribution approximation in formula \eqref{eq:p-asy}, the upper bound, which is decided by \eqref{equ: upper bound}, can be written as 
\begin{equation*}
 m_L - \Phi^{-1}(\rho_u) \sqrt{m_L(1-p)} - (1 + \delta_u)\alpha L = 0,   
\end{equation*}
For any $p \in[0, 1]$, we have
\begin{equation*}
  m_L - \Phi^{-1}(\rho_u) \sqrt{m_L} - (1 + \delta_u)\alpha L \le 0,   
\end{equation*}
which gives the upper bound of $m_L$ as 
\begin{equation}
  m_L 
  \le
  u(\alpha, L) \triangleq
  \left\lceil
  \left(
    \sqrt{\frac{1}{4}\Phi(\rho_u)^2 +  (1 + \delta_u)\alpha L} + \frac{1}{2}\Phi(\rho_u) 
  \right)^2 \right\rceil,
\end{equation}
where $\lceil \cdot \rceil$  is the ceiling function.
With a similar argument, we can obtain the lower bound of $m_L$ as
\begin{equation}
  m_L 
  \ge 
  l(\alpha, L) 
  \triangleq 
  \left\lfloor 
  \left(
    \sqrt{\frac{1}{4}\Phi(\rho_l)^2 + (1-\delta_l)\alpha L} - \frac{1}{2}\Phi(\rho_l) 
  \right)^2 
  \right\rfloor,
\end{equation}
where $\lfloor \cdot \rfloor$ is the floor function. 
If the $m_L$ is smaller than the lower bound $l(\alpha, L)$, or larger than the upper bound $u(\alpha, L)$, we can stop the adaptive procedure and draw inference using estimated p-value $\hat{p}_L$. 
Pseudo code of our proposed adaptive procedure is summarized in Algorithm \ref{alg: adaptive method}.

\begin{algorithm}
\caption{Pseudo code of our proposed adaptive method}
\label{alg: adaptive method}
\begin{algorithmic}[1]
    \State \textbf{Input:} $\alpha, u(\alpha, L), l(\alpha, L)$
    \State \textbf{Output:} $\hat p_L$  
    \State \textbf{Initialization:} $L = 1000$ and $\mathcal L_{\max}$ \Comment{$\mathcal L_{\max}$ can be set by formula \eqref{equ: PR formula} following the PR re-randomization method \citep{plamadeala2012sequential}}
    
    \While{$L \leq \mathcal L_{\max}$}  
        \State calculate $m_L$
        \If{$m_L < l(\alpha, L)$ or $m_L > u(\alpha, L)$}
          \State estimate $\hat{p}_L$ and draw inference  based on the nominal significance bound $\alpha$. 
        \Else
          \State $L = L + 1000$
        \EndIf
    \EndWhile 
\end{algorithmic}
\end{algorithm}

An example to demonstrate our adaptive procedure is provided as follows. 
Here, we set $\delta_u = \delta_l = 0.1$ and $\rho_u = \rho_L = 0.99$ in \eqref{equ: upper bound} and \eqref{equ: lower bound}.
Furthermore, we assume the nominal significance bound $\alpha = 0.0001$, and we set $L_{\max} = 6,636,000$ according to the PR re-randomization \cite{plamadeala2012sequential}. 
The rule table for the adaptive procedure is shown in Table \ref{tbl: adaptive rule table with alpha 0.0001}.
We start with $L = 1,000$ repetitions and observe 3 replications with equal or more extreme value than the observed value, i.e., $m_{L = 1,000} = 3$.
Clearly, 3 falls in the range formed by $l(\alpha, L)= 0$ and $u(\alpha, L) = 6$, so we move to the next $L = 2,000$.
For $L = 2,000$, if $m_{L = 2,000} = 7 \notin [0, 6]$, 
we can stop the iteration and draw statistical inference at $\alpha = 0.0001$.
The estimated $\hat p_{L = 2,000} = 7/2,000$.

\begin{table*}[htbp]
    \caption{Lower and Upper bound of the adaptive procedure at $\alpha = 0.0001$
    \label{tbl: adaptive rule table with alpha 0.0001}}
    \centering
    \begin{adjustbox}{max width=0.9\textwidth}
    \centering
    \begin{threeparttable}
    \begin{tabular}{p{0.12\textwidth}p{0.15\textwidth}p{0.15\textwidth}p{0.2\textwidth}p{0.2\textwidth}}
        \hline
        $L$ & $l(\alpha, L)$ & $u(\alpha, L)$ & $p_l$ & $p_u$ \\ 
        \hline
        1,000     &    0     &     6    & 0.000000 & 0.006000 \\ 
        2,000     &    0     &     6    & 0.000000 & 0.003000 \\ 
        3,000     &    0     &     7    & 0.000000 & 0.002333 \\ 
        4,000     &    0     &     7    & 0.000000 & 0.001750 \\ 
        5,000     &    0     &     7    & 0.000000 & 0.001400 \\ 
        $\vdots$  & $\vdots$ & $\vdots$ & $\vdots$ & $\vdots$ \\
        10,000    &     0    &     8    & 0.000000 & 0.000800 \\ 
        $\vdots$  & $\vdots$ & $\vdots$ & $\vdots$ & $\vdots$ \\
        50,000    &     1    &    15    & 0.000020 & 0.000300 \\ 
        $\vdots$  & $\vdots$ & $\vdots$ & $\vdots$ & $\vdots$ \\
        100,000   &     4    &    22    & 0.000040 & 0.000220\\
        $\vdots$  & $\vdots$ & $\vdots$ & $\vdots$ & $\vdots$ \\
        500,000   &    31    &    76    & 0.000062 & 0.000152 \\
        $\vdots$  & $\vdots$ & $\vdots$ & $\vdots$ & $\vdots$ \\
        1,000,000 &    70    &   138    & 0.000070 & 0.000138 \\
        $\vdots$  & $\vdots$ & $\vdots$ & $\vdots$ & $\vdots$ \\
        2,000,000 &   151    &   258    & 0.000076 & 0.000129 \\
        $\vdots$  & $\vdots$ & $\vdots$ & $\vdots$ & $\vdots$ \\
        3,000,000 &   234    &   376    & 0.000078 & 0.000125 \\
        $\vdots$  & $\vdots$ & $\vdots$ & $\vdots$ & $\vdots$ \\
        4,000,000 &   318    &   492    & 0.000080 & 0.000123 \\
        $\vdots$  & $\vdots$ & $\vdots$ & $\vdots$ & $\vdots$ \\
        5,000,000 &   403    &   608    & 0.000081 & 0.000122 \\
        $\vdots$  & $\vdots$ & $\vdots$ & $\vdots$ & $\vdots$ \\
        6,636,000 &   543    &   796    & 0.000082 & 0.000120 \\ 
        \hline
    \end{tabular}
    \begin{tablenotes}
      \footnotesize
       \item[1] This table is generated by setting $\delta_u = \delta_l = 0.1$ and $\rho_u = \rho_L = 0.99$ in \eqref{equ: upper bound} and \eqref{equ: lower bound}.
       \item[2] $p_l = l(\alpha, L)/L$ and $p_u = u(\alpha, L)/L$ are the lower and upper bound of the observed p-value.
    \end{tablenotes}
    \end{threeparttable}
    \end{adjustbox}
\end{table*}

\section{Simulations}
\label{sec: simulation}
In this section, we evaluate the numerical performance of the three methods described in the last section for the re-randomization test in three scenarios: (1) continuous outcome with Wald test, (2) binary outcome with Fisher's exact test, and (3) time-to-event outcome with the log-rank test.
For our proposed adaptive procedure, we set $\delta_u = \delta_l = 0.1$ and $\rho_u = \rho_L = 0.99$ in \eqref{equ: upper bound} and \eqref{equ: lower bound}.
We will first introduce the data generation mechanism in Section \ref{sec: simulation - data generation}.
Then, we will present the simulation results in Section \ref{sec: simulation - result}.

\subsection{Data Generation Mechanism}
\label{sec: simulation - data generation}

In this section, we discuss the data generation mechanism for three scenarios. 
First, we assume the sample size is 200 with two treatment groups. Second, we assume there are 3 stratification factors, 50 sites, 2 ECOGs, and 3 TMBs.
Given these three stratification factors, the 200 subjects are classified into two groups by minimization randomization \cite[see][]{jin2019algorithms}.
Equal weights are used in each stratification factor for minimization randomization, with an allocation probability of 0.9 for the imbalanced group. 
Third, different coefficients are assigned to the stratification factors, namely,
\begin{equation*}
    \left\{
    \begin{array}{l}
        \beta_{\text{site}} 
        = 
        (
          \beta_{\text{site}, 1}, 
          \ldots, 
          \beta_{\text{site}, 50}
        )^{\top} 
        \text{ with } \beta_{\text{site}, j} \overset{i.i.d.}{\sim} N(0.2, 1), \; j = 1, \ldots, 50\\
        \beta_{\text{ecog}} = (0.3, 0.5)^\top \\
        \beta_{\text{tmb}} = (-0.3, 0, 0.3)^\top 
    \end{array}
    \right..
\end{equation*}
Based on these coefficients, one can generate continuous, binary, and survival outcomes from linear predictors. 
Specifically, for the $i$-th subject from $j_1$-th site, $j_2$-th ECOG, and $j_3$-th TMB and $j_4$-th group, the linear predictor for the $i$-th subject is
$$
  l_i = \beta_{\text{site}, j_1} + \beta_{\text{ecog}, j_2} + \beta_{\text{tmb}, j_3} + \beta_{\text{group},j_4}, 
$$ 
where
$\beta_{\text{site}, j_1}, \beta_{\text{ecog}, j_2}, 
\beta_{\text{tmb}, j_3}, \beta_{\text{group}, j_4}$ 
indicating the $j_1, j_2, j_3, j_4$-th elements in the vector 
$\beta_{\text{site}}$, $\beta_{\text{ecog}}$, 
$\beta_{\text{tmb}}$, $\beta_{\text{group}}$, respectively. 
For $\beta_{\text{group}}$, 
it is selected to ensure proper p-value in each case based on 1 million repetitions. 
The outcome model specification is given in Section \ref{sec: simulation - continous} - Section \ref{sec: simulation - survival}.

We assume the significance level is fixed as $\alpha = 0.01$ for all three scenarios.
We further assume there is a treatment effect, whose corresponding p-value is 
$
  p 
  \in 
  \{
    0.005, 0.007, 
$
$
  0.008, 0.009, 0.011, 0.013, 0.015, 0.020 
  \}.
$
For each combination of $\alpha$ and $p$,
we apply three methods to determine the number of repetitions for the re-randomization test: 
(1) parametric re-randomization 
(2) PR re-randomization, and
(3) adaptive procedure.
These three methods will be applied to 1,000 simulated datasets from continuous, binary or survival outcomes. 
And we set $\mathcal L_{\max} = 66,000$ for the maximum number of repetitions in adaptive procedure with the reference from the PR re-randomization method. 
We evaluate the performance by the metrics in Table \ref{tbl: simulation metrics}.

\begin{table}[htbp]
    \caption{Metrics used in simulations \label{tbl: simulation metrics}}
    \centering
    \begin{adjustbox}{max width=\textwidth}
    \centering
    \footnotesize
    \begin{threeparttable}
    \begin{tabular}{p{0.05\textwidth}p{0.27\textwidth}p{0.3\textwidth}p{0.3\textwidth}}
      \hline
      Metrics & Definition  &  Mathematical formula\tnote{1}  & Criterion \\
      \hline
      $\overline{L}$ & The average number of repetitions  
      & $\overline{L} = \frac{1}{N_{\text{sim}}} \left(L^{(1)} + L^{(2)} + \ldots + L^{(N_{\text{sim}})}\right)$  
      & \pbox{5cm}{The smaller value of $\overline{L}$, the shorter computational time. \\} \\
      $L_{\max}$ 
      & The maximum number of repetitions
      & $L_{\max} = \max \left(L^{(1)} + L^{(2)} + \ldots + L^{(N_{\text{sim}})}\right)$ 
      &\pbox{5cm}{The smaller value of $L_{\max}$, the shorter computational time. \\}\\
      $L_{\min}$ 
      & The minimal number of repetitions\textsuperscript{2}
      & $L_{\min} = \min \left(L^{(1)} + L^{(2)} + \ldots + L^{(N_{\text{sim}})}\right)$
      & \pbox{5cm}{The smaller value of $L_{\min}$, the shorter computational time.\\ }\\
      $P_{\max}(\%)$ 
      & \pbox{5cm}{The proportion of simulations to reach  $\mathcal L_{\max}$. \textsuperscript{3}} 
      & $P_{\max}(\%) = \frac{1}{N_{\text{sim}}}  \sum_{i=1}^{N_{\text{sim}}} \mathbbm 1\{L^{(i)} = \mathcal L_{\max}\}$
      & \pbox{5cm}{The smaller value of $P_{\max}(\%)$, the shorter computational time.\\ }\\
      \text{CP}($\%$) 
      & \pbox{5cm}{The concordance percentage that $\hat p_\infty$ and $\hat p_L$ draw consistent inference conclusions when the significance level $\alpha = 0.01$.\textsuperscript{4} }
      & $\text{CP}($\%$) = \frac{1}{N_{\text{sim}}}  \sum_{i=1}^{N_{\text{sim}}} \mathbbm 1\{\text{$\hat p_\infty$, $\hat p_{L^{(i)}}$ are consistent}\}$
      & \pbox{5cm}{The larger value of \text{CP}($\%$), the better performance in correct statistical inference. }\\
      \hline
    \end{tabular}
    \begin{tablenotes}
      \footnotesize
      \item[1] $N_{\text{sim}}$ is the total number of simulations. In our simulation, we set $N_{\text{sim}} = 1000$. And the notation $L^{(i)}$ is the number of repetitions in the $i$-th simulation for $i = 1, \ldots, N_{\text{sim}}$.
      \item[2] For the adaptive procedure, the least value is 1,000.
      \item[3] Only applicable to our proposed adaptive procedure.
      \item[4] Here we use the estimated p-value based on 1 million repetitions to approximate $\hat{p}_\infty$, i.e., $\hat{p}_\infty \approx \hat{p}_{1000,000}$.
    \end{tablenotes}
    \end{threeparttable}
    \end{adjustbox}
\end{table}

\subsubsection{Continuous Outcome}
\label{sec: simulation - continous}
For the continuous outcome, we generate the outcome $y_i$ by
$$
  y_i 
  =  
  l_i +
  e_i,
$$
where $e_i \overset{i.i.d.}{\sim} N(0, 1)$ is from the standard normal distribution.

For each instance of randomization, the linear regression model is performed. 
The estimated p-value under $L$ repetitions (denoted as $\hat p_L$) is the number of the fraction of randomization with the value of the treatment effect coefficient as extreme or more extreme based on the Wald test statistics.

\subsubsection{Binary Outcome}
\label{sec: simulation - binary}

For the binary outcome, the response is the random variable from a binomial distribution with probability of success as the inverse logit transformation of the coefficients.
$$
  y_i \overset{i.i.d.}{\sim} \text{Binomial}(\tilde p_i),
$$
where $\tilde p_i = \exp(l_i)/(1 + \exp(l_i))$.

For each instance of randomization, the logistic regression is performed. 
The estimated p-value under $L$ repetitions (denoted as $\hat p_L$) is the number of the fraction of randomization with the value of the treatment effect coefficient as extreme or more extreme based on the Wald test statistics.

\subsubsection{Time-to-event Outcome}
\label{sec: simulation - survival}
For the time-to-event outcome, the survival time is generated from a proportional hazard (PH) model with uniform censoring with a censoring proportion around 50\%.
Mathematically speaking, its survival time and survival status are 
$$
  \left\{
  \begin{array}{l}
    y_i = \min\{T_i, C_i\} \\
    \delta_i = \mathbbm 1\{T_i < C_i \},
  \end{array}
  \right.
$$
where $T_i = \exp(l_i + \log(-\log(u_i))$ is the response for the $i$-th subject with $u_i \overset{i.i.d.}{\sim} \text{uniform}(0, 1)$, and $C_i \overset{i.i.d.}{\sim} \text{uniform}(0, 5)$ is the censoring time for the $i$-th subject.
Besides, $\mathbbm 1\{\cdot\}$ is the indicator function, i.e., $\delta_i = 1$ if $T_i < C_i$, otherwise $\delta_i = 0$.

For each instance of randomization, the stratified log-rank test is performed.
The stratum considered include ECOGs, and TMBs. 
The p-value of the re-randomization test is equal to the fraction
of randomizations with the value of the log-rank statistics as extreme or more extreme as the one obtained from the stratified log-rank test of the actual data.

\subsection{Simulation Results}
\label{sec: simulation - result}
The performance of the three approaches under three types of outcomes is summarized in Table \ref{tbl: simulation - continous} - Table \ref{tbl: simulation - survival}.

For the PR re-randomization method (i.e., M1 in Section \ref{sec: review of existing methods}), we use $L = 65,695$ repetitions across three scenarios given the fixed $\alpha=0.01$.
We find it has robust and good performance under different types of outcomes and different values of $p$, i.e., its CP(\%) always stays above 99\%. 

For the parametric re-randomization method (i.e., M2 in Section \ref{sec: review of existing methods}), we investigate three number of repetitions, i.e., $L = 1,000, 5,000$, and $10,000$. 
We observe its performance improves after increasing the number of repetitions in all outcome types.
For example, under the continuous endpoint with $p = 0.008$, its CP(\%) is 92.2\%, 99.9\% and 100.0\% when $L = 1,000, 5,000$, and $10,000$, respectively.
Since a higher value of CP(\%) indicates better performance, we find performance of parametric re-randomization method improves when $L$ increases.
The case with 10,000 repetitions has good performance for the continuous and survival outcome, given its high CP(\%) values. 
For the binary outcome, when $p = 0.009$ and $\alpha = 0.01$, its CP(\%) is only $67.4\%$ even under $10,000$ repetitions. 
One potential reason is due to the heavy tail of the null distribution estimated from the re-randomization test that is different from a normal distribution. 
The deviation of a normal distribution is expected for a covariate-adaptive randomization procedure, given the dependency between samples.
Thus, using a normal approximation for a parametric re-randomization procedure can be misleading. 
More theoretical investigations are required on the asymptotic distribution of test statistics under covariate-adaptive re-randomization, which is beyond the scope of the current paper.

\begin{figure}[htbp]
	\centering
	\begin{tabular}{ccc}
		\includegraphics[width = 0.3\textwidth]{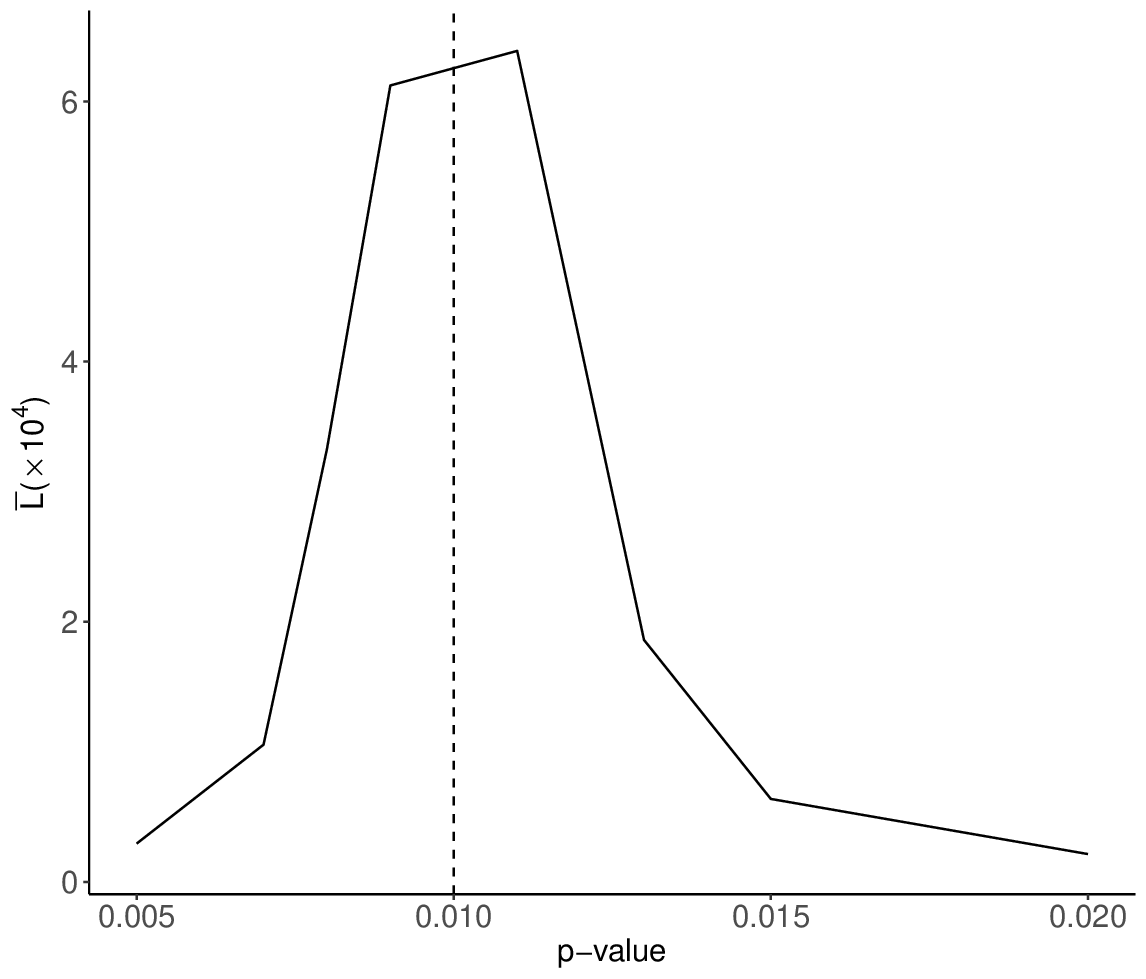} &
		\includegraphics[width = 0.3\textwidth]{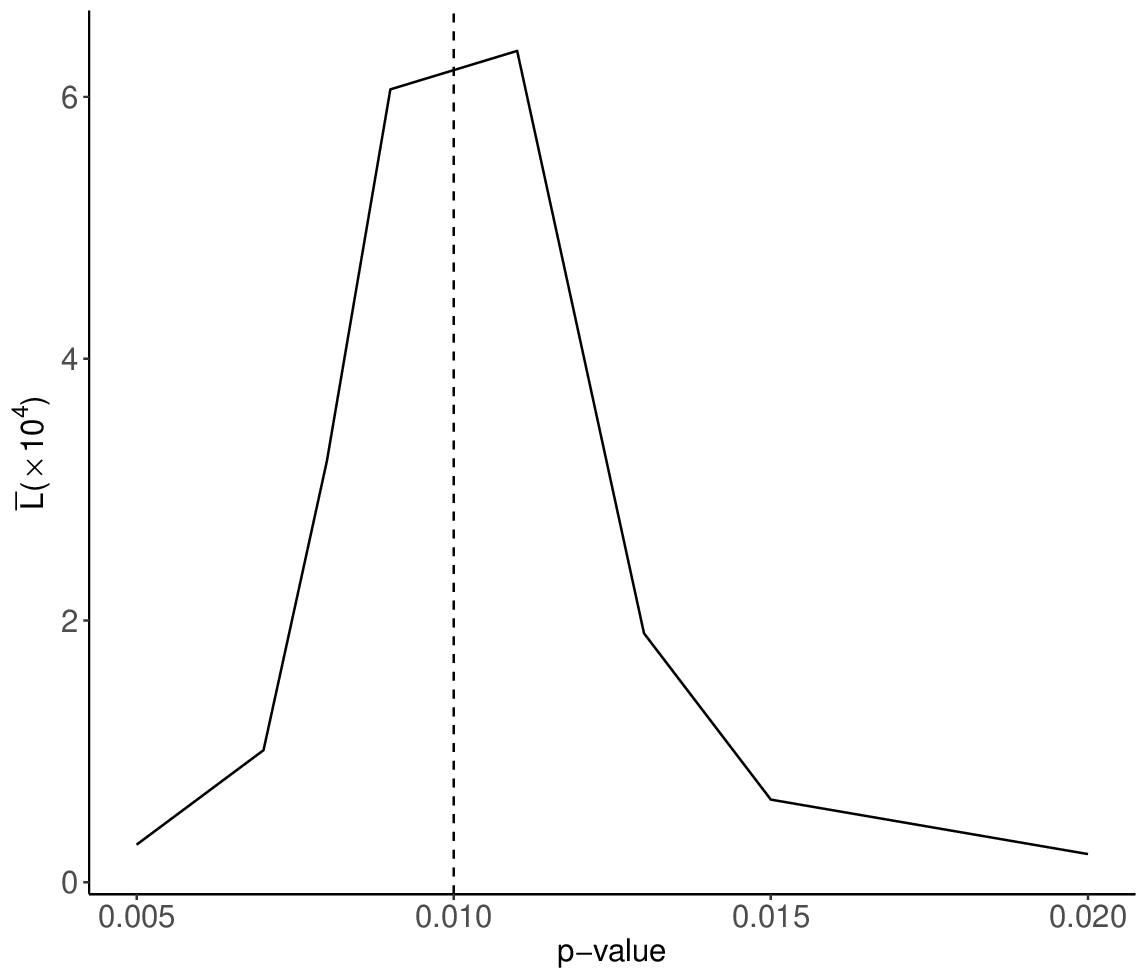} &
		\includegraphics[width = 0.3\textwidth]{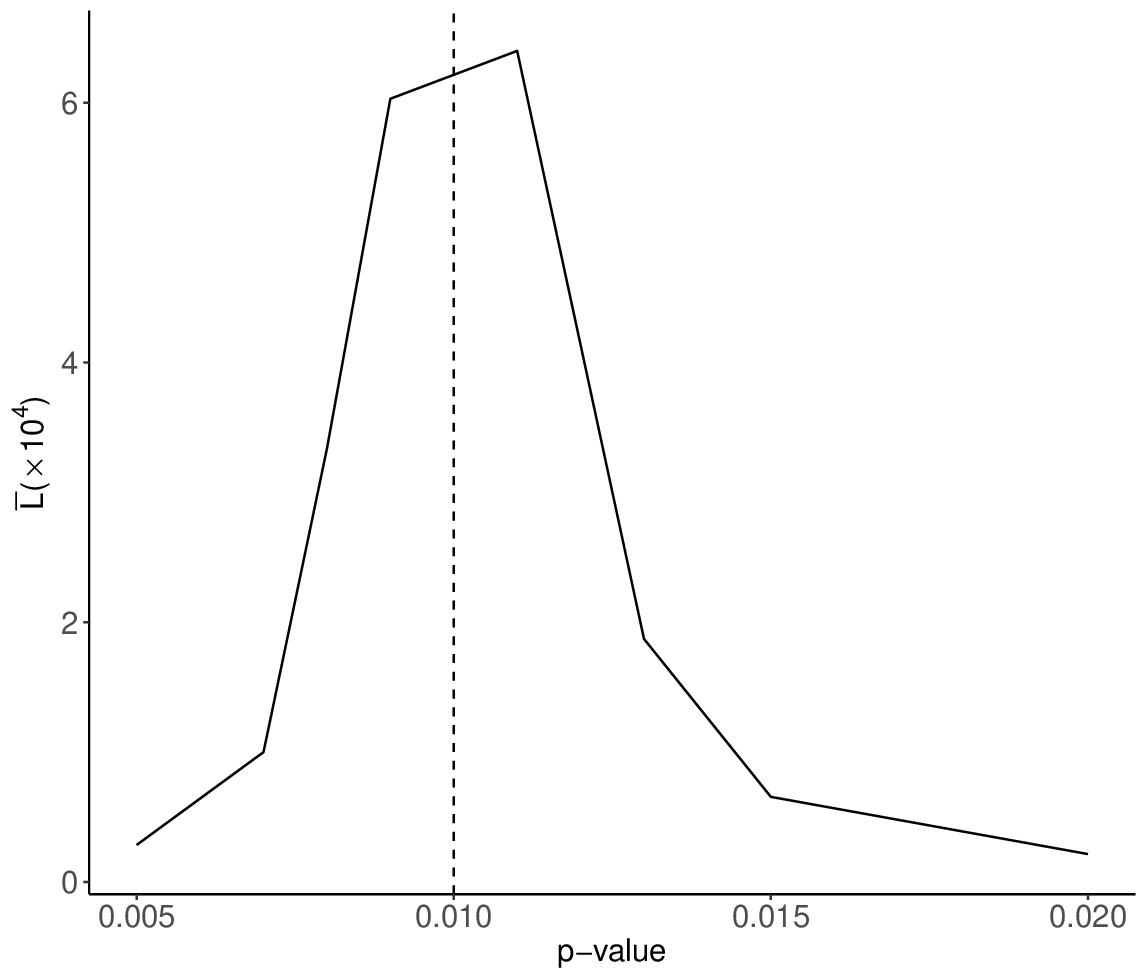} \\
		(a) continuous &  (b) binary & (c) survival \\
	\end{tabular}
	\caption{Average number of repetitions of the adaptive procedure under different types of endpoints \label{fig: adp L curve}}
\end{figure}

For the adaptive procedure method, the number of repetitions varies under different scenarios and $p$. 
And we visualize the trend of the repetitions under the different values of $p$ in Figure \ref{fig: adp L curve}.
From this figure, we find, overall, the adaptive procedure has a much smaller average number of repetitions when $p$ is far away from $\alpha$ compared with the PR re-randomization method (i.e., $L = 65, 695$). 
For example, the adaptive procedure only requires an average of only $2,854$ repetitions when $p = 0.005$ under the survival endpoints, while it takes more repetitions when $p$ is close to $\alpha$ as expected. 
Accordingly, more repetitions are required to ensure statistical efficiency. 
Additionally, its CP(\%) value is near or above 99\% across different $p$ and types of outcomes.

By comparing the above three methods, we find the adaptive procedure and PR re-randomization are better than parametric re-randomization, especially when $p$ is close to $\alpha$ and under binary/survival outcome.
For example, for the binary outcome, when $p = 0.009$ and $\alpha = 0.01$, there is only around $67.4\%$ chance for the parametric re-randomization method to make consistent inferences as infinite repetitions.
Both the adaptive procedure and PR re-randomization have close or above $99\%$ to make consistent inferences.
The adaptive procedure requires a much smaller average number of repetitions compared with the PR re-randomization, especially when $p$ is far away from $\alpha$.
For example, under the survival outcome and $p = 0.005$, the adaptive procedure only takes an average of approximately 3,000 repetitions.
While the PR re-randomization needs 65,695 repetitions.

\begin{table}[htbp]
    \caption{Summary of Simulation Results on Continuous Endpoints when $\alpha = 0.01$ \textsuperscript{1}
    \label{tbl: simulation - continous}}
    \centering
    \begin{adjustbox}{max width=\textwidth}
    \centering
    \footnotesize
    \begin{threeparttable}
    \begin{tabular}{p{0.2\textwidth}p{0.13\textwidth}p{0.13\textwidth}p{0.13\textwidth}p{0.1\textwidth}p{0.1\textwidth}}
        \hline
        Method 
        & $\overline{L}$ 
        & $ L_{\max}$ 
        & $ L_{\min}$ 
        & $P_{\max}(\%)$  
        & CP($\%$)   \\ 
        \hline
        \multicolumn{6}{c}{$p$ = 0.005} \\ 
        \hline
        parametric-1000 &  1,000 &  1,000 &  1,000 & - & $100.00$ \\ 
        parametric-5000 &  5,000 &  5,000 &  5,000 & - & $100.00$ \\ 
        parametric-10000 & 10,000 & 10,000 & 10,000 & - & $100.00$ \\ 
        PR & 65,695 & 65,695 & 65,695 & - & $100.00$ \\ 
        adaptive procedure &  2,948 & 12,000 &  1,000 & 0 & $100.00$ \\ 
        \hline
        \multicolumn{6}{c}{$p$ = 0.007} \\ 
        \hline
        parametric-1000 &  1,000 &  1,000 &  1,000 & - & $97.80$ \\ 
        parametric-5000 &  5,000 &  5,000 &  5,000 & - & $100.00$ \\ 
        parametric-10000 & 10,000 & 10,000 & 10,000 & - & $100.00$ \\ 
        PR & 65,695 & 65,695 & 65,695 & - & $100.00$ \\ 
        adaptive procedure & 10,551 & 57,000 &  1,000 & 0 & $100.00$ \\ 
        \hline
        \multicolumn{6}{c}{$p$ = 0.008} \\ 
        \hline
        parametric-1000 &  1,000 &  1,000 &  1,000 & - & $92.20$ \\ 
        parametric-5000 &  5,000 &  5,000 &  5,000 & - & $99.90$ \\ 
        parametric-10000 & 10,000 & 10,000 & 10,000 & - & $99.90$ \\ 
        PR & 65,695 & 65,695 & 65,695 & - & $100.00$ \\ 
        adaptive procedure & 33,269 & 66,000 &  1,000 & 18.6 & $100.00$ \\ 
        \hline
        \multicolumn{6}{c}{$p$ = 0.009} \\ 
        \hline
        parametric-1000 &  1,000 &  1,000 &  1,000 & - & $73.00$ \\ 
        parametric-5000 &  5,000 &  5,000 &  5,000 & - & $94.70$ \\ 
        parametric-10000 & 10,000 & 10,000 & 10,000 & - & $98.70$ \\ 
        PR & 65,695 & 65,695 & 65,695 & - & $99.80$ \\ 
        adaptive procedure & 61,228 & 66,000 &  1,000 & 91.2 & $99.80$ \\ 
        \hline
        \multicolumn{6}{c}{$p$ = 0.011} \\ 
        \hline
        parametric-1000 &  1,000 &  1,000 &  1,000 & - & $70.90$ \\ 
        parametric-5000 &  5,000 &  5,000 &  5,000 & - & $89.50$ \\ 
        parametric-10000 & 10,000 & 10,000 & 10,000 & - & $96.60$ \\ 
        PR & 65,695 & 65,695 & 65,695 & - & $99.80$ \\ 
        adaptive procedure & 63,885 & 66,000 &  1,000 & 95.6 & $99.10$ \\ 
        \hline
        \multicolumn{6}{c}{$p$ = 0.013} \\ 
        \hline
        parametric-1000 &  1,000 &  1,000 &  1,000 & - & $93.20$ \\ 
        parametric-5000 &  5,000 &  5,000 &  5,000 & - & $100.00$ \\ 
        parametric-10000 & 10,000 & 10,000 & 10,000 & - & $100.00$ \\ 
        PR & 65,695 & 65,695 & 65,695 & - & $100.00$ \\ 
        adaptive procedure & 18,605 & 66,000 &  1,000 & 0.4 & $99.80$ \\ 
        \hline
        \multicolumn{6}{c}{$p$ = 0.015} \\ 
        \hline
        parametric-1000 &  1,000 &  1,000 &  1,000 & - & $99.10$ \\ 
        parametric-5000 &  5,000 &  5,000 &  5,000 & - & $100.00$ \\ 
        parametric-10000 & 10,000 & 10,000 & 10,000 & - & $100.00$ \\ 
        PR & 65,695 & 65,695 & 65,695 & - & $100.00$ \\ 
        adaptive procedure &  6,380 & 26,000 &  1,000 & 0.0 & $100.00$ \\ 
        \hline
        \multicolumn{6}{c}{$p$ = 0.020} \\ 
        \hline
        parametric-1000 &  1,000 &  1,000 &  1,000 & - & $100.00$ \\ 
        parametric-5000 &  5,000 &  5,000 &  5,000 & - & $100.00$ \\ 
        parametric-10000 & 10,000 & 10,000 & 10,000 & - & $100.00$ \\ 
        PR & 65,695 & 65,695 & 65,695 & - & $100.00$ \\ 
        adaptive procedure &  2,142 &  5,000 &  1,000 & 0.0 & $100.00$ \\
        \hline
    \end{tabular}
    \begin{tablenotes}
      \footnotesize
      \item[1] The upper limit is of the adaptive procedure is 66,000  based on PR re-ranomization.
    \end{tablenotes}
    \end{threeparttable}
    \end{adjustbox}
\end{table}

\begin{table}[htbp]
    \caption{Summary of Simulation Results on Binary Endpoints when $\alpha = 0.01$ \textsuperscript{1}
    \label{tbl: simulation - binary}}
    \centering
    \begin{adjustbox}{max width=\textwidth}
    \centering
    \begin{threeparttable}
    \footnotesize
    \begin{tabular}{p{0.2\textwidth}p{0.13\textwidth}p{0.13\textwidth}p{0.13\textwidth}p{0.1\textwidth}p{0.1\textwidth}}
        \hline
        Method 
        & $\overline{L}$ 
        & $ L_{\max}$ 
        & $ L_{\min}$ 
        & $P_{\max}(\%)$  
        & CP($\%$)   \\ 
        \hline
        \multicolumn{6}{c}{$p$ = 0.005} \\ 
        \hline
        parametric-1000 &  1,000 &  1,000 &  1,000 & - & $99.90$ \\ 
        parametric-5000 &  5,000 &  5,000 &  5,000 & - & $100.00$ \\ 
        parametric-10000 & 10,000 & 10,000 & 10,000 & - & $100.00$ \\ 
        PR & 65,695 & 65,695 & 65,695 & - & $100.00$ \\ 
        adaptive procedure &  2,871 & 13,000 &  1,000 & 0.0 & $100.00$ \\ 
        \hline
        \multicolumn{6}{c}{$p$ = 0.007} \\ 
        \hline
        parametric-1000 &  1,000 &  1,000 &  1,000 & - & $92.80$ \\ 
        parametric-5000 &  5,000 &  5,000 &  5,000 & - & $99.40$ \\ 
        parametric-10000 & 10,000 & 10,000 & 10,000 & - & $99.80$ \\ 
        PR & 65,695 & 65,695 & 65,695 & - & $100.00$ \\ 
        adaptive procedure & 10,083 & 51,000 &  1,000 & 0.0 & $100.00$ \\ 
        \hline
        \multicolumn{6}{c}{$p$ = 0.008} \\ 
        \hline
        parametric-1000 &  1,000 &  1,000 &  1,000 & - & $78.00$ \\ 
        parametric-5000 &  5,000 &  5,000 &  5,000 & - & $91.40$ \\ 
        parametric-10000 & 10,000 & 10,000 & 10,000 & - & $93.90$ \\ 
        PR & 65,695 & 65,695 & 65,695 & - & $100.00$ \\ 
        adaptive procedure & 32,196 & 66,000 &  1,000 & 17.5 & $100.00$ \\ 
        \hline
        \multicolumn{6}{c}{$p$ = 0.009} \\ 
        \hline
        parametric-1000 &  1,000 &  1,000 &  1,000 & - & $56.90$ \\ 
        parametric-5000 &  5,000 &  5,000 &  5,000 & - & $64.70$ \\ 
        parametric-10000 & 10,000 & 10,000 & 10,000 & - & $67.40$ \\ 
        PR & 65,695 & 65,695 & 65,695 & - & $99.80$ \\ 
        adaptive procedure & 60,572 & 66,000 &  1,000 & 89.7 & $99.80$ \\ 
        \hline
        \multicolumn{6}{c}{$p$ = 0.011} \\ 
        \hline
        parametric-1000 &  1,000 &  1,000 &  1,000 & - & $81.80$ \\ 
        parametric-5000 &  5,000 &  5,000 &  5,000 & - & $90.30$ \\ 
        parametric-10000 & 10,000 & 10,000 & 10,000 & - & $93.40$ \\ 
        PR & 65,695 & 65,695 & 65,695 & - & $99.30$ \\ 
        adaptive procedure & 63,513 & 66,000 &  1,000 & 94.7 & $98.60$ \\ 
        \hline
        \multicolumn{6}{c}{$p$ = 0.013} \\ 
        \hline
        parametric-1000 &  1,000 &  1,000 &  1,000 & - & $95.10$ \\ 
        parametric-5000 &  5,000 &  5,000 &  5,000 & - & $99.70$ \\ 
        parametric-10000 & 10,000 & 10,000 & 10,000 & - & $100.00$ \\ 
        PR & 65,695 & 65,695 & 65,695 & - & $100.00$ \\ 
        adaptive procedure & 19,016 & 66,000 &  1,000 & 0.6 & $99.80$ \\ 
        \hline
        \multicolumn{6}{c}{$p$ = 0.015} \\ 
        \hline
        parametric-1000 &  1,000 &  1,000 &  1,000 & - & $99.00$ \\ 
        parametric-5000 &  5,000 &  5,000 &  5,000 & - & $100.00$ \\ 
        parametric-10000 & 10,000 & 10,000 & 10,000 & - & $100.00$ \\ 
        PR & 65,695 & 65,695 & 65,695 & - & $100.00$ \\ 
        adaptive procedure &  6,309 & 25,000 &  1,000 & 0.0 & $100.00$ \\ 
        \hline
        \multicolumn{6}{c}{$p$ = 0.020} \\ 
        \hline
        parametric-1000 &  1,000 &  1,000 &  1,000 & - & $100.00$ \\ 
        parametric-5000 &  5,000 &  5,000 &  5,000 & - & $100.00$ \\ 
        parametric-10000 & 10,000 & 10,000 & 10,000 & - & $100.00$ \\ 
        PR & 65,695 & 65,695 & 65,695 & - & $100.00$ \\ 
        adaptive procedure &  2,148 &  7,000 &  1,000 & 0.0 & $100.00$ \\ 
        \hline
    \end{tabular}
    \begin{tablenotes}
      \footnotesize
      \item[1] The upper limit is of the adaptive procedure is 66,000  based on PR re-ranomization.
    \end{tablenotes}
    \end{threeparttable}
    \end{adjustbox}
\end{table}

\begin{table}[htbp]
    \caption{Summary of Simulation Results on Survival Endpoints when $\alpha = 0.01$ \textsuperscript{1}
    \label{tbl: simulation - survival}}
    \centering
    \begin{adjustbox}{max width=\textwidth}
    \centering
    \begin{threeparttable}
    \footnotesize
    \begin{tabular}{p{0.2\textwidth}p{0.13\textwidth}p{0.13\textwidth}p{0.13\textwidth}p{0.1\textwidth}p{0.1\textwidth}}
        \hline
        Method 
        & $\overline{L}$ 
        & $ L_{\max}$ 
        & $ L_{\min}$ 
        & $P_{\max}(\%)$  
        & CP($\%$)   \\ 
        \hline
        \multicolumn{6}{c}{$p$ = 0.005} \\ 
        \hline
        parametric-1000 &  1,000 &  1,000 &  1,000 & - & $100.00$ \\ 
        parametric-5000 &  5,000 &  5,000 &  5,000 & - & $100.00$ \\ 
        parametric-10000 & 10,000 & 10,000 & 10,000 & - & $100.00$ \\ 
        PR & 65,695 & 65,695 & 65,695 & - & $100.00$ \\ 
        adaptive procedure &  2,854 & 10,000 &  1,000 & 0.0 & $100.00$ \\ 
        \hline
        \multicolumn{6}{c}{$p$ = 0.007} \\ 
        \hline
        parametric-1000 &  1,000 &  1,000 &  1,000 & - & $99.30$ \\ 
        parametric-5000 &  5,000 &  5,000 &  5,000 & - & $100.00$ \\ 
        parametric-10000 & 10,000 & 10,000 & 10,000 & - & $100.00$ \\ 
        PR & 65,695 & 65,695 & 65,695 & - & $100.00$ \\ 
        adaptive procedure & 10,002 & 46,000 &  1,000 & 0.0 & $100.00$ \\ 
        \hline
        \multicolumn{6}{c}{$p$ = 0.008} \\ 
        \hline
        parametric-1000 &  1,000 &  1,000 &  1,000 & - & $97.10$ \\ 
        parametric-5000 &  5,000 &  5,000 &  5,000 & - & $99.80$ \\ 
        parametric-10000 & 10,000 & 10,000 & 10,000 & - & $100.00$ \\ 
        PR & 65,695 & 65,695 & 65,695 & - & $100.00$ \\ 
        adaptive procedure & 33,371 & 66,000 &  1,000 & 19.3 & $100.00$ \\ 
        \hline
        \multicolumn{6}{c}{$p$ = 0.009} \\ 
        \hline
        parametric-1000 &  1,000 &  1,000 &  1,000 & - & $90.80$ \\ 
        parametric-5000 &  5,000 &  5,000 &  5,000 & - & $96.30$ \\ 
        parametric-10000 & 10,000 & 10,000 & 10,000 & - & $98.80$ \\ 
        PR & 65,695 & 65,695 & 65,695 & - & $100.00$ \\ 
        adaptive procedure & 60,312 & 66,000 &  1,000 & 89.3 & $100.00$ \\ 
        \hline
        \multicolumn{6}{c}{$p$ = 0.011} \\ 
        \hline
        parametric-1000 &  1,000 &  1,000 &  1,000 & - & $75.60$ \\ 
        parametric-5000 &  5,000 &  5,000 &  5,000 & - & $93.50$ \\ 
        parametric-10000 & 10,000 & 10,000 & 10,000 & - & $98.30$ \\ 
        PR & 65,695 & 65,695 & 65,695 & - & $99.70$ \\ 
        adaptive procedure & 63,999 & 66,000 &  1,000 & 95.6 & $99.20$ \\ 
        \hline
        \multicolumn{6}{c}{$p$ = 0.013} \\ 
        \hline
        parametric-1000 &  1,000 &  1,000 &  1,000 & - & $95.80$ \\ 
        parametric-5000 &  5,000 &  5,000 &  5,000 & - & $100.00$ \\ 
        parametric-10000 & 10,000 & 10,000 & 10,000 & - & $100.00$ \\ 
        PR & 65,695 & 65,695 & 65,695 & - & $100.00$ \\ 
        adaptive procedure & 18,720 & 66,000 &  1,000 & 0.5 & $99.90$ \\ 
        \hline
        \multicolumn{6}{c}{$p$ = 0.015} \\ 
        \hline
        parametric-1000 &  1,000 &  1,000 &  1,000 & - & $99.80$ \\ 
        parametric-5000 &  5,000 &  5,000 &  5,000 & - & $100.00$ \\ 
        parametric-10000 & 10,000 & 10,000 & 10,000 & - & $100.00$ \\ 
        PR & 65,695 & 65,695 & 65,695 & - & $100.00$ \\ 
        adaptive procedure &  6,559 & 24,000 &  1,000 & 0.0 & $100.00$ \\ 
        \hline
        \multicolumn{6}{c}{$p$ = 0.020} \\ 
        \hline
        parametric-1000 &  1,000 &  1,000 &  1,000 & - & $100.00$ \\ 
        parametric-5000 &  5,000 &  5,000 &  5,000 & - & $100.00$ \\ 
        parametric-10000 & 10,000 & 10,000 & 10,000 & - & $100.00$ \\ 
        PR & 65,695 & 65,695 & 65,695 & - & $100.00$ \\ 
        adaptive procedure &  2,153 &  7,000 &  1,000 & 0.0 & $100.00$ \\ 
        \hline
    \end{tabular}
    \begin{tablenotes}
      \footnotesize
      \item[1] The upper limit is of the adaptive procedure is 66,000  based on PR re-ranomization.
    \end{tablenotes}
    \end{threeparttable}
    \end{adjustbox}
\end{table}

\section{Computation Time}
\label{sec: computation time}
Based on the simulation results in the last section, 
the proposed adaptive procedure can significantly reduce
the average number of repetitions when the estimated p-value is 
relatively larger or smaller than the significance level $\alpha$.
(i.e. $\hat{p}_\infty < 0.9 \alpha$ or $\hat{p}_\infty > 1.1 \alpha$).
However, it is still possible to require a huge number of repetitions if the estimated p-value is close to a tiny $\alpha$. 
Without loss of generality, we provide strategies to reduce total computation time by assuming we need to replicate there-randomization procedure 1 million times. 

Strategy 1: improve re-randomization algorithm implementation. 
We implement the minimization algorithm in R \citep{zhao2022electronic} by carefully avoiding repeating computations in calculating the imbalance of randomization \citep{zhao2022survey}. 
For 1,000 subjects, we compare our implementation with the `Minirand' package version 0.1.3 available on CRAN \citep{jin2019algorithms}. 
Based on the example in the `Minirand` function, our implementation completes one re-randomization in 0.84 seconds vs. ``Minirand'' in 5.38 seconds using a computer with i5-10310U CPU \@1.70GHz 2.21GHz.
The R code is provided in the Appendix. 
The speed can be further improved if we implemented the algorithm in C++ using the ``Rcpp'' package \citep{eddelbuettel2011rcpp}, which is future work. 

Strategy 2: parallel computing. 
With our improved re-randomization algorithm implementation,
it can still take around 5.6 days to complete 1 million repetitions procedure to assign a treatment group for 1,000 subjects. 
The total analysis time for the 1 million repetitions can be even longer. 
We can reduce the total computation time by using off-the-shelf 
R packages that support parallel execution. 
For example, the “foreach” package can easily be applied to execute the re-randomization procedure in multiple cores of a CPU or multiple nodes of a high-performance computing (HPC) cluster \citep{calaway2015package}.
For an HPC cluster with hundreds of processors, the total computation time can be reduced to multiple hours instead of multiple days.  
An R code example of expediting the computation by the ``foreach'' R package is available in the Appendix.
Other examples of using ``foreach'' in the literature can be found in \cite{kane2013scalable} and \cite{king2015statistical}.

\section{Discussion}
\label{sec: discussion}

The re-randomization test is performed 
by fixing the order of randomization of the study subjects, 
their covariates, and/or responses. 
Then the randomization procedure is repeated many times. We suggest an adaptive procedure to reduce the average number of repetitions for the re-randomization test.  

For an ongoing clinical trial before a database lock, 
baseline covariates are commonly available for the blinded study team.
For covariate-adaptive randomization like 
minimization procedure blinded study team
can prepare a set of (e.g., one million) replications of treatment assignment based on pre-specified randomization procedure before the database lock.
This step can be completed after all
the baseline covariates are collected for 
all randomized subjects.   
The generated treatment group can also be reused for different endpoints (e.g. PFS and OS) in primary and sensitivity analyses. 
After the database lock, the study team 
only needs to validate that the covariates 
are the same before they perform the analysis. 
If the response can be finalized before the database lock, we can further complete the replicated analyses 
based on the generated treatment group before the database lock.
In this way, the study team can save valuable time 
after the database lock.

\newpage
\section*{Appendix}
\subsection*{Hypothetical group sequential design.}

We consider a group sequential design with an initial one-sided $\alpha$-level at 0.005 for progression-free survival (PFS). Table \ref{tbl: gs summary} shows the boundary properties for 
one-sided $\alpha$-level of the interim analysis and final analysis.
The p-value boundaries are derived using a Lan-DeMets O’Brien-Fleming spending function. 
Here, "gsDesign" 3.2.2 is used for study design.

Based on the design, the study will randomize 734 participants in a 1:1 ratio into the experimental or control group. 
Based on the expected number of 467 events at the final analysis and one interim analysis at approximately 50\% of the target number of events, the study has approximately 90\% power at an overall alpha level of 0.5\% (1-sided) if the true hazard ratio is 0.70. 

The sample size  calculations for PFS and the derivation of the projected times for the interim and final analyses assume the following:

\begin{itemize}
    \item There is one interim analysis and one final analysis.
    \item The hazard ratio for the PFS endpoint is a constant, and we assume PFS follows an exponential distribution with a median of 6 months for the control group. 
    \item The monthly drop-out rate is 0.01 for PFS. 
    \item The enrollment period is 6 months, and the study duration is 18 months with 6 months as minimum follow-up duration.
\end{itemize}

Table \ref{tbl: gs summary} summarizes the operational characteristics of the hypothetical group sequential design. 

\begin{table}[htbp]
    \caption{Summary of the Hypothetical Group Sequential Design\label{tbl: gs summary}}
    \centering
    \begin{adjustbox}{max width=\textwidth}
    \centering
    \begin{threeparttable}
    \begin{tabular}{p{0.25\textwidth}p{0.35\textwidth}p{0.3\textwidth}}
        \hline
        Analysis & Value & Efficacy \\ 
        \hline
        Interim Analysis 1:  & Z & 3.801369 \\ 
        N: 644 
        & p (1-sided)\textsuperscript{1} & 0.000072 \\ 
        Events: 234 
        & $\sim$ HR at bound\textsuperscript{2} & 0.607911 \\
        Month: 10.5 
        & P(Cross) if HR=1\textsuperscript{3} & 0.000072 \\ 
        & P(Cross) if HR=0.7\textsuperscript{4} & 0.141649 \\
        \hline
        Final Analysis & Z & 2.577571 \\ 
        N: 734 
        & p (1-sided) & 0.004975 \\ 
        Events: 467 
        & $\sim$ HR at bound & 0.787697 \\ 
        Month: 18 
        & P(Cross) if HR=1 & 0.005000 \\ 
        & P(Cross) if HR=0.7 & 0.900000 \\ 
        \hline
    \end{tabular}
    \begin{tablenotes}
      \footnotesize
      \item[1]p(1-sided) is the nominal alpha for testing.
      \item[2]\textasciitilde{}HR at bound is the approximated HR required to reach an efficacy bound.
      \item[3]P(Cross) if HR = 1 is the probability of crossing bound under the null hypothesis. 
      \item[4]P(Cross) if HR = 0.7 is the probability of crossing bound under the alternative hypothesis. 
    \end{tablenotes}
    \end{threeparttable}
    \end{adjustbox}
    
\end{table}

\pagebreak

\subsection*{R code for group sequential design.}

\begin{verbatim}
library(gsDesign)

gsSurv(
  # parameters related to sample size
  alpha = 0.005,            # type I error
  beta = 0.1,               # type II error
  ratio = 1,                # randomization ratio (experimental : control)
  # parameters related to timing
  k = 2,                    # number of analyses (interim + final)
  # parameters related to boundaries
  test.type = 1,            # one-sided test
  sfu = sfLDOF,             # spending function of upper bound 
  sfl = sfLDOF,             # spending function of lower bound    
  # parameters related to hazards
  lambdaC = log(2)/6,       # hazard rates for the control group
  hr = 0.7,                 # hazard ratio (experimental : control)
  hr0 = 1,                  # hazard ratio under H1
  # parameters related to failure rates
  eta = 0.01,               # dropout hazard rates for the control group
  # parameters related to enrollment rates
  gamma = 10,               # rates of entry
  R = 6,                    # duration of time periods for recruitment rates
  T = 18,                   # study duration
  minfup = 6                # minimum follow-up duration
  ) 
\end{verbatim}

\subsection*{R code for minimization randomization}
\label{sec: appendix - computation time comparision}

In this section, we provide code examples to compare 
our implementation with the 'Minirand' package version 0.1.3 available on CRAN \cite[see][]{jin2019algorithms}.
We take 1,000 subjects as an example.

The implementation code of our minimization algorithm is presented below.

\begin{verbatim}
library(rerandom)

set.seed(1234)
n <- 1000                       # sample size 

# assume equal probability in each stratum
stratum <- define_stratum(
  gender = c("1" = 0.4, "0" = 0.6), 
  age = c("1" = 0.3, "0" = 0.7),
  hypertension = c("2" = 0.33, "1" = 0.2, "0" = 0.5), 
  use_antibiotics = c("1" = 0.33, "0" = 0.67))

treatment <- c("group1", "group2", "group3")


df <- simu_stratum(n = n, stratum = stratum) 
df <- simu_treatment_minimization(
  df = df, 
  treatment = treatment, 
  prob = define_prob(0.9, treatment), 
  ratio = c(2, 2, 1),         # randomization ratio
  imbalance_fun = imbalance_fun_range)}
\end{verbatim}

The implementation of 'Minirand' package is presented below.

\begin{verbatim}
library(Minirand)
set.seed(1234)
ntrt <- 3                       # number of arms
nsample <- 1000                 # sample size
trtseq <- c(1, 2, 3)            # arm index
ratio <- c(2, 2, 1)             # randomization ratio

# 4 stratum, similar to 
# gender, age, hypertension, use_antibiotics in our implementation
c1 <- sample(seq(1, 0), nsample, replace = TRUE, prob = c(0.4, 0.6)) 
c2 <- sample(seq(1, 0), nsample, replace = TRUE, prob = c(0.3, 0.7))
c3 <- sample(c(2, 1, 0), nsample, replace = TRUE, prob = c(0.33, 0.2, 0.5)) 
c4 <- sample(seq(1, 0), nsample, replace = TRUE, prob = c(0.33, 0.67)) 
# generate the matrix of covariate factors for the subjects
covmat <- cbind(c1, c2, c3, c4) 
# label of the covariates 
colnames(covmat) = c("Gender", "Age", "Hypertension", "Use of Antibiotics") 
# equal weights
covwt <- c(1/4, 1/4, 1/4, 1/4) 
# result is the treatment needed from minimization method
res <- rep(100, nsample) 

# generate treatment assignment for the 1st subject

res[1] = sample(trtseq, 1, replace = TRUE, prob = ratio/sum(ratio))
for (j in 2:nsample)
{
  # get treatment assignment sequential for all subjects
  res[j] <- Minirand(
    covmat = covmat, j, covwt = covwt, ratio = ratio, 
    ntrt = ntrt, trtseq = trtseq, method = "Range", 
    result = res, p = 0.9)
}
trt1 <- res

\end{verbatim}

\subsection*{Parallel Computing}
\label{sec: appendix - parallel computation}

We provide code examples to enable parallel computation for minimization randomization. 

\begin{verbatim}
library(rerandom)
library(dplyr)
library(foreach)

# register the parallel backend 
registerDoParallel(8)

df <- list()
foreach(task_id = 1:1000) %dopar% {
    set.seed(task_id)
    
    #---------- Simulation Setup --------------------#
    
    n <- 200
    n_rerandom <- 1e+6
    
    stratum <- define_stratum(
      site = rep(1, 50),
      ecog = c("0" = 1, "1" = 1),
      tmb = c("<=6" = 1, ">6 and <= 12" = 1, ">12" = 1) )
    
    treatment <- c("drug", "placebo")
    
    df[[i]] <- n %>% 
      simu_stratum(stratum = stratum) %>%
      simu_treatment_minimization(treatment = treatment, 
                                  prob = define_prob(0.9, treatment), 
                                  ratio = c(1, 1),
                                  imbalance_fun = imbalance_fun_range)
}    
\end{verbatim}

\bibliography{reference}%

\clearpage



\end{document}